\RequirePackage{fix-cm}
\documentclass[smallextended]{svjour3}       
\smartqed  
\usepackage{graphicx}
\begin{document}

\title{Quantum teleportation, entanglement,
and Bell nonlocality in Unruh channel
}


\author{Soroush Haseli         \and
}


\institute{Soroush Haseli \at
              Faculty of Physics, Urmia University of Technology, Urmia, Iran \\
              \email{soroush.haseli@uut.ac.ir}           
}

\date{Received: date / Accepted: date}

\maketitle

\begin{abstract}
Decoherence is an unavoidable phenomenon that results from the interaction of the system with its surroundings.  The study of decoherence due to the  relativistic effects has the fundamental importance. The Unruh effect is observed by the relativistically accelerator observer. The unruh effect acts as a quantum channel and we call it the Unruh Channel. The Unruh channel can be characterize  by providing its Kraus representation.  We consider the bipartite scheme in which the quantum information is shared between an inertial observer (Alice) and an accelerated observer (Rob) in the case of Dirac field. We will show that this channel reduces the common quantum information between the two observers. In this work we will study the effects of the Unruh channel on various facets of quantum correlations, such as the quantum teleportation, entanglement, and Bell inequality violations for a Dirac field mode.
\end{abstract}

\section{Introduction}\label{intro}
 Decoherence is an unavoidable phenomenon that results from the interaction of the system with its surroundings. As a result of the interaction of the system with its surroundings quantum correlation, as a fundamental resource for quantum information processing, decreases. The study of decoherence due to the  relativistic effects has the fundamental importance both from  a fundamental perspective as well as to assist in future experiments involving a relativistic observers. The relativistic effect, known as the Unruh effect \cite{1,2,3}, states that from the point of view of the accelerating moving observer with acceleration $a$ the Minkowski vacuum appears as a hot gas emitting the radiation of the black-body at Unruh temperature  
 \begin{equation}
 \tau=\frac{\hbar a}{2 \pi k_B c}
 \end{equation}
 \begin{figure}[h]
  \centering
  \includegraphics[width=0.31\textwidth]{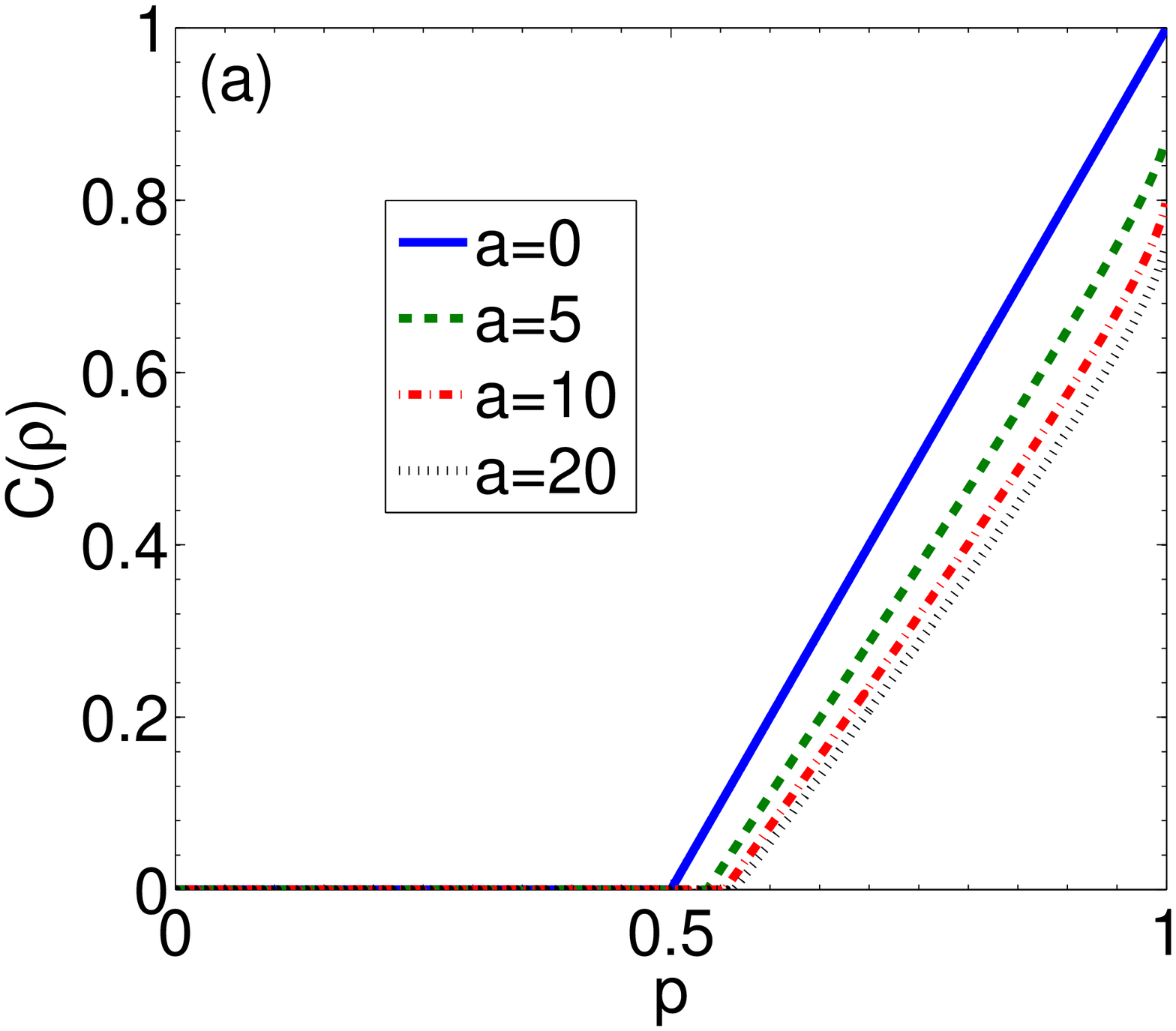}\quad
  \includegraphics[width=0.31\textwidth]{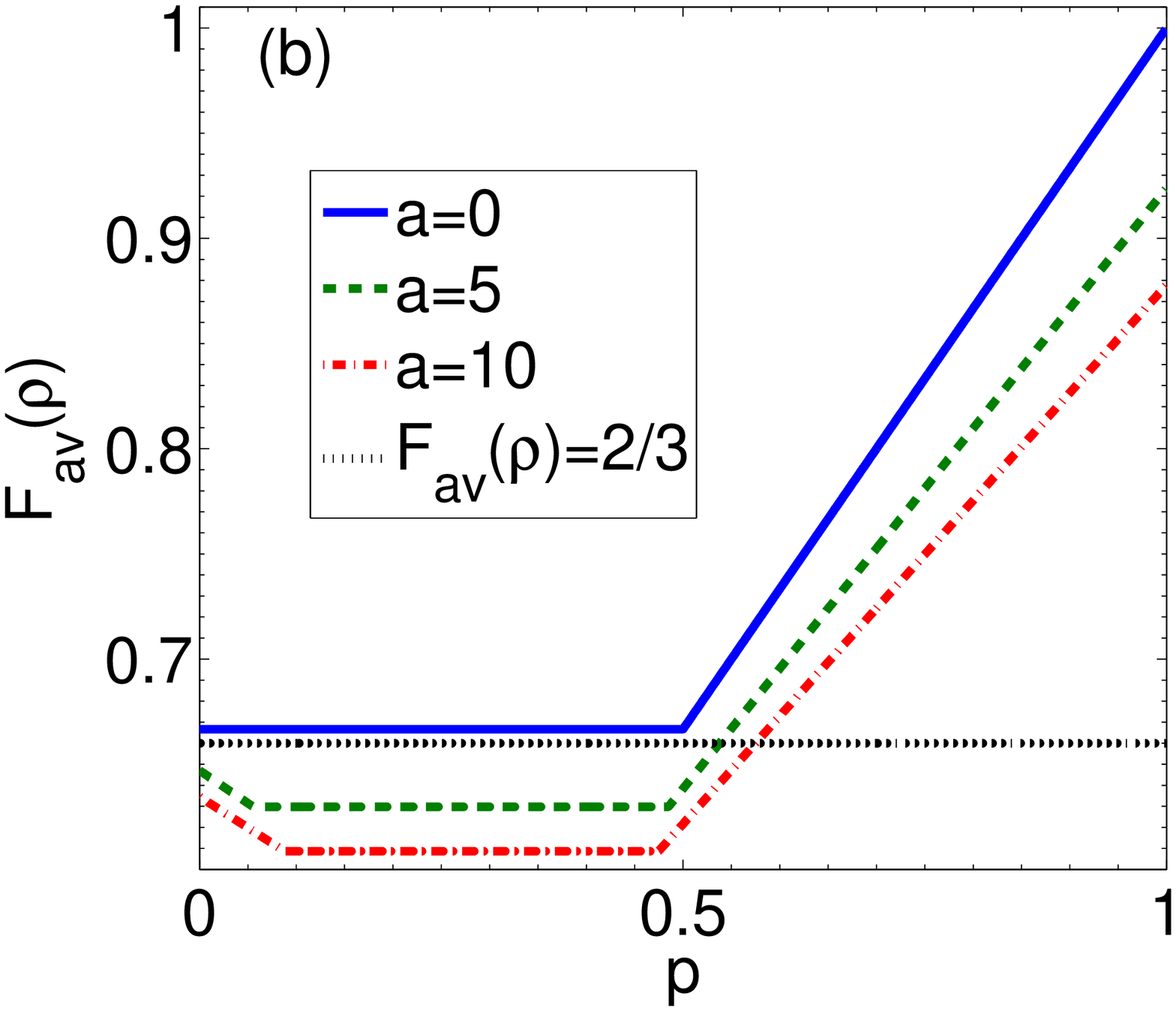}\quad
  \includegraphics[width=0.31\textwidth]{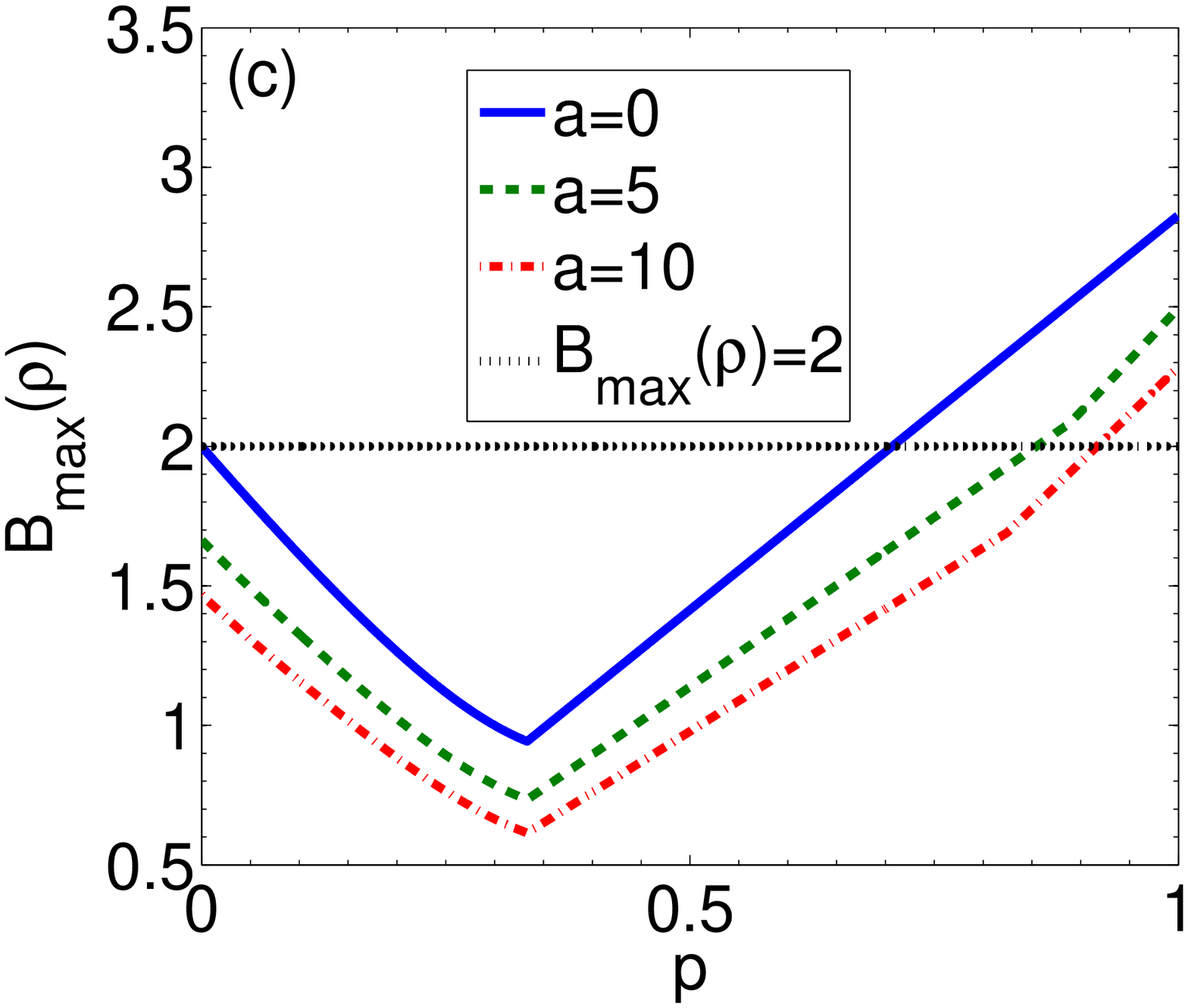}
\caption{(a) Concurrence. (b). Teleportation fidelity (c). Bell non-locality as a function of probability parameter for different value of acceleration $a$, when Alice and Rob initially share the Bell diagonal state $\rho^{AR}=p \vert \psi^- \rangle\langle \psi^- \vert + \frac{1-p}{2}(\vert \psi^+ \rangle\langle \psi^+ \vert + \vert \phi^+ \rangle\langle \phi^+ \vert)$, and $\omega=1$.}
\label{figure1}
\end{figure}
\begin{figure}[h]
  \centering
  \includegraphics[width=0.31\textwidth]{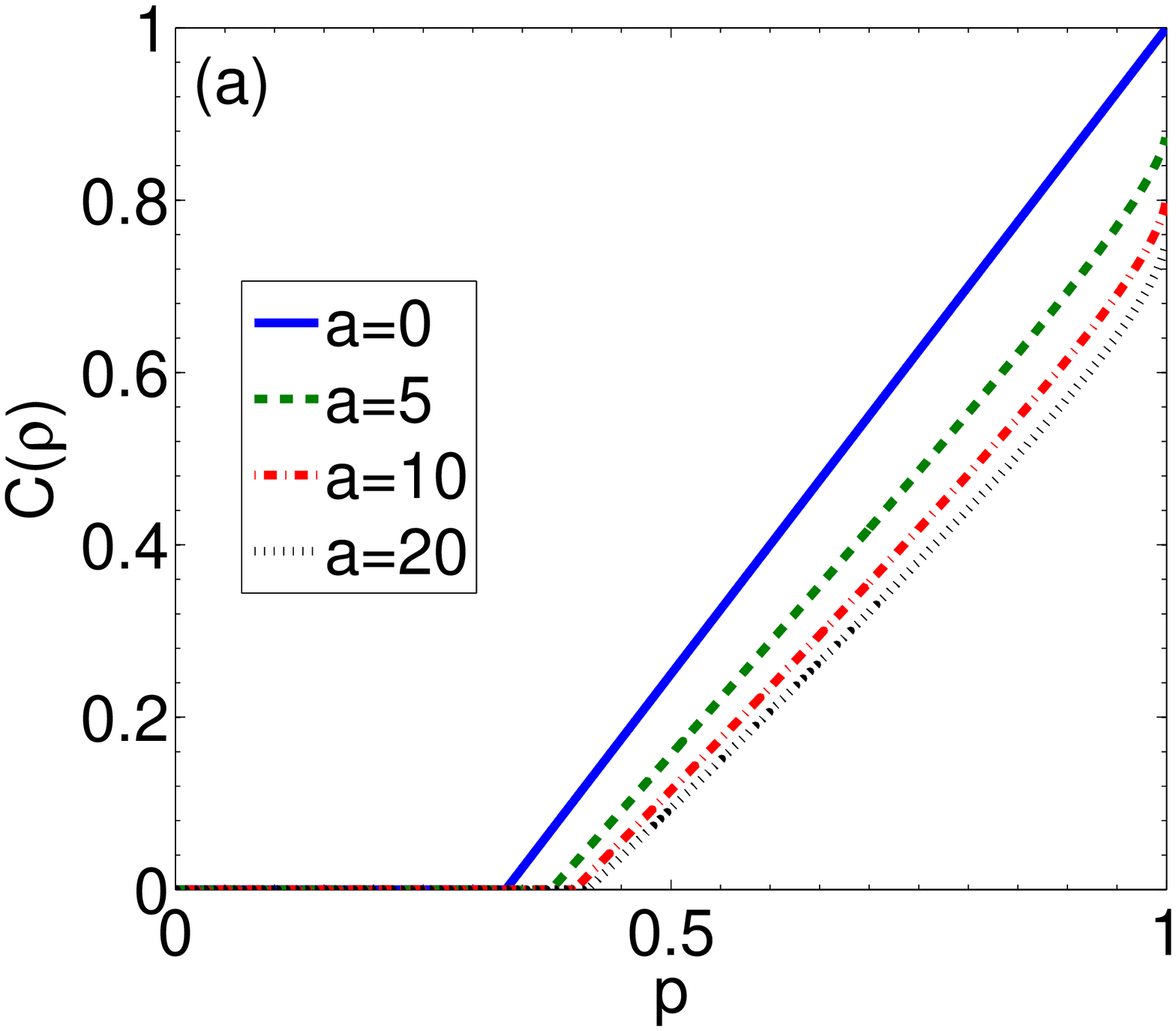}\quad
  \includegraphics[width=0.31\textwidth]{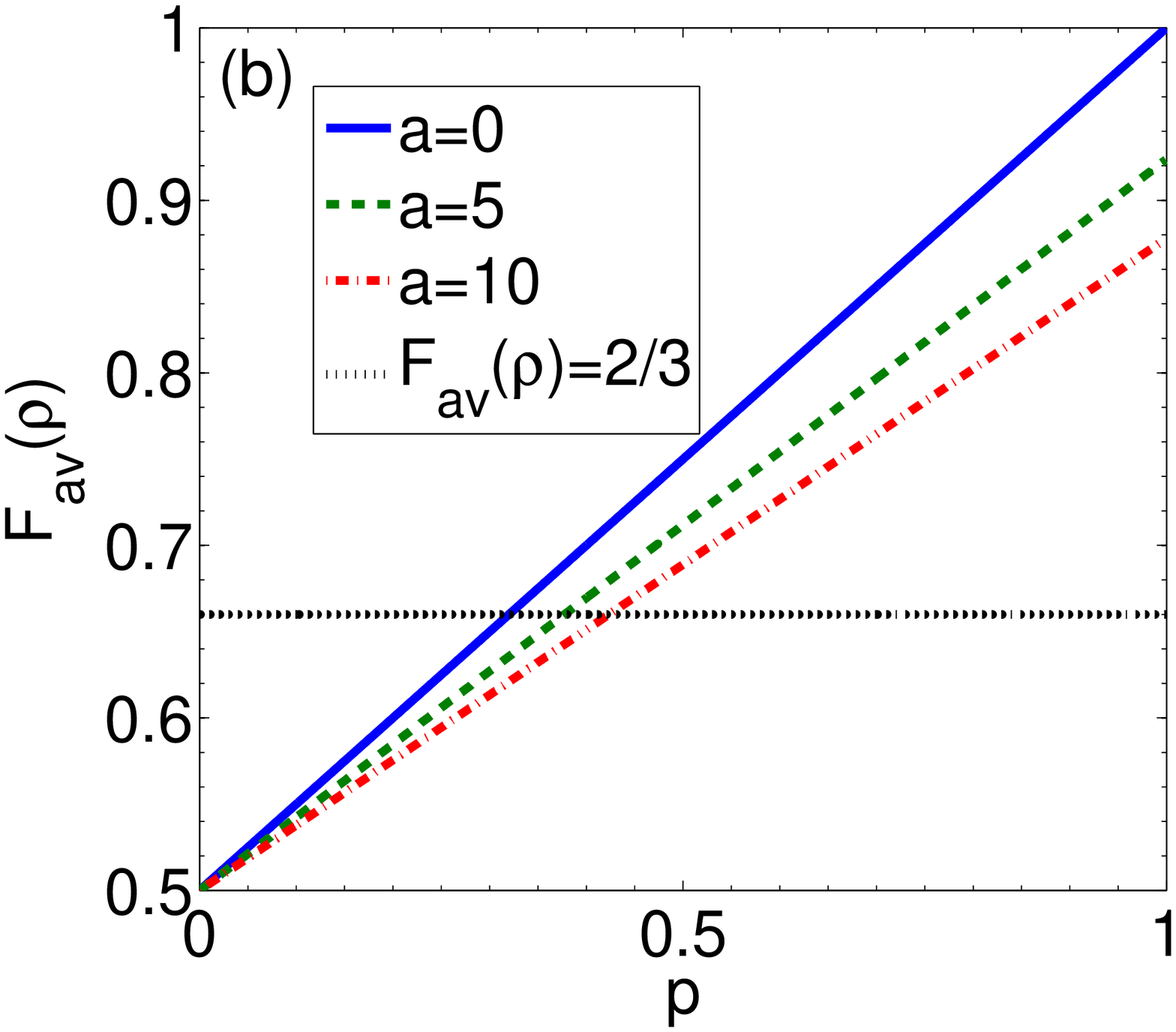}\quad
  \includegraphics[width=0.31\textwidth]{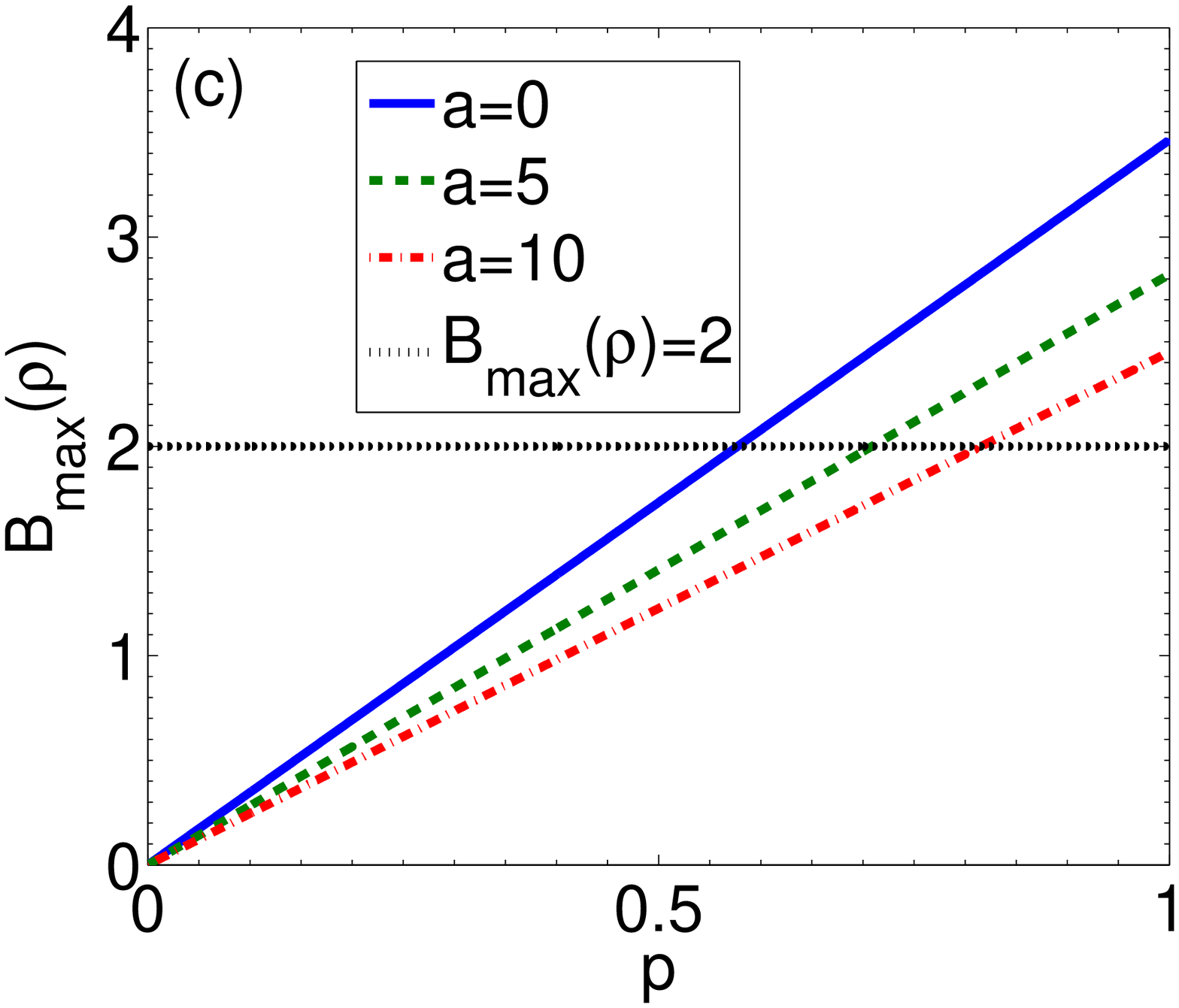}
\caption{(a) Concurrence. (b). Teleportation fidelity (c). Bell non-locality as a function of probability parameter for different value of acceleration $a$, when Alice and Rob initially share a two-qubit werner state $\rho^{AR}=\frac{1-p}{4}I \otimes I + p \vert \psi^{-} \rangle \langle \psi^{-} \vert$, and $\omega=1$.}
\label{figure2}
\end{figure}
\begin{figure}[h]
  \centering
  \includegraphics[width=0.31\textwidth]{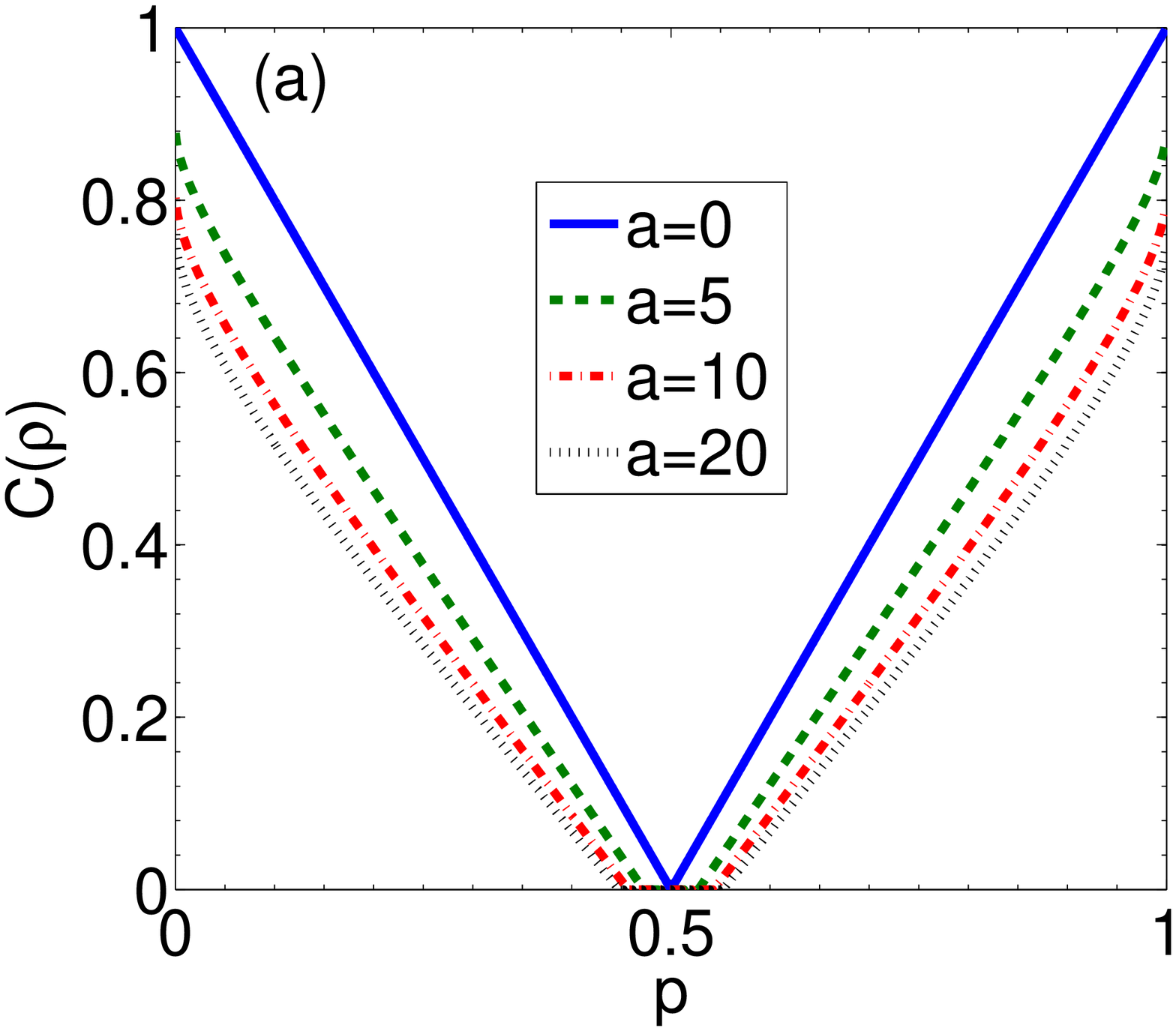}\quad
  \includegraphics[width=0.31\textwidth]{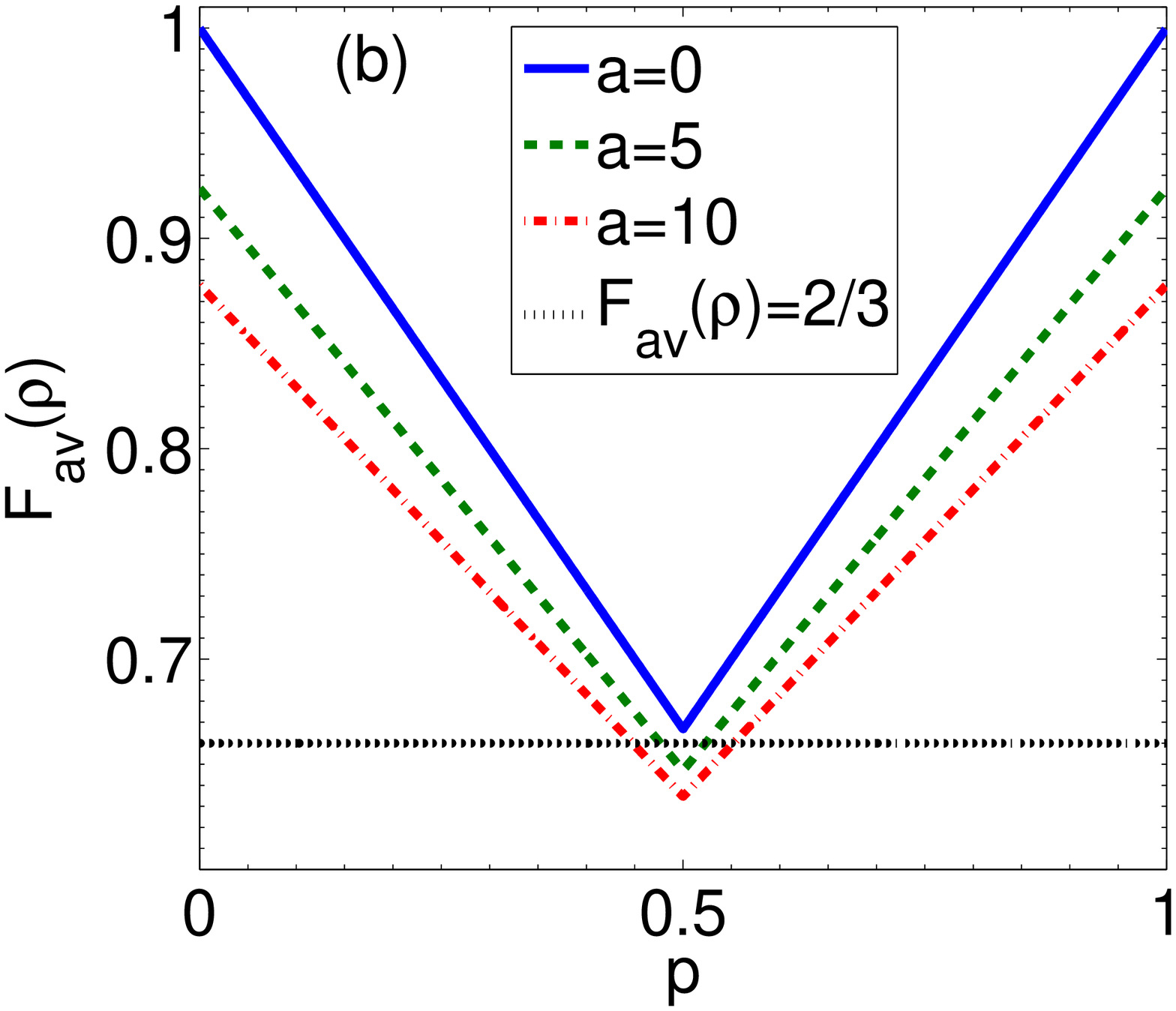}\quad
  \includegraphics[width=0.31\textwidth]{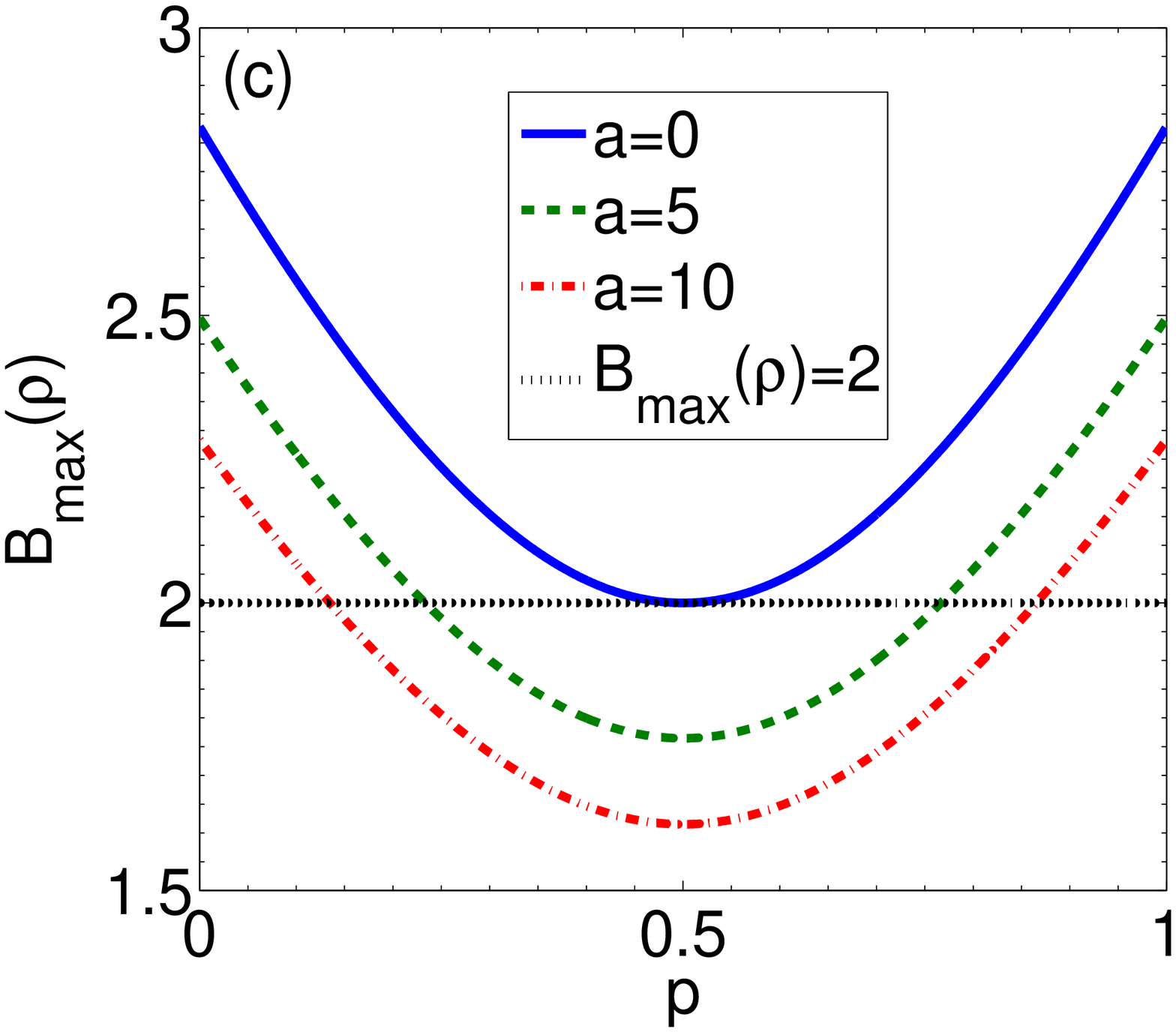}
\caption{(a) Concurrence. (b). Teleportation fidelity (c). Bell non-locality as a function of probability parameter for different value of acceleration $a$, when Alice and Rob initially share a two-qubit X-state $\rho^{AR}=(1-p) \vert \phi^{+} \rangle \langle \phi^{+} \vert + p \vert \psi^{+} \rangle \langle \psi^{+} \vert,$, and $\omega=1$.}
\label{figure3}
\end{figure}
 where $c$ is the speed of light in vacuum and $k_B$ is Boltzmann's constant.   Unruh effect is calculated with an idealization where acceleration continues  for infinite time. Of course, there have been some interesting studies exploring the finite acceleration aspects of Unruh effect  \cite{4}. The role of interaction time of detectors with the quantum field on Unruh effect has also been studied \cite{5,6}. The Unruh effect creates a decoherence-like effect \cite{7}.  It reduces the quantum information shared between an inertial observer (Alice) and an accelerated observer (Rob), as seen in the latter’s frame, in the case of bosonic or Dirac field modes \cite{8,9}. The study of Unruh effect is part of efforts to understand the relativistic concepts of quantum information theory\cite{10,11,12,13,14,15}. The Unruh channel has the particular importance from the point of view of quantum information theory because it is conjugate degradable channel\cite{16}. A channel is conjugate degradable when the environment can be simulated from the output of the channel.  This feature of the Unruh channel makes it possible to calculate the classical and quantum capacity of the channel, as well as the exchange between these two capacities for a number of scenarios \cite{17,18}. The most important step towards the experimental realization in the field of relativistic quantum information can be obtained through circuit quantum electrodynamics, using superconducting quantum interferometric devices \cite{19}. The concept of geometric phase can also be used to suggest a possible detection of Unruh temperature at sufficiently small accelerations that are experimentally available \cite{20}. In this work, by providing a geometric characterization of the Unruh channel, we will study various types of quantum correlations, such as, Bell inequality violations, entanglement and teleportation  under the effect of the Unruh
channel.  Things that have been done so far in the context of Unruh channel are consist of either the  fermionic \cite{21} or bosonic \cite{22} channels, depending on whether one is working with a Dirac or a scalar field , respectively. According to the finite occupation of the fermionic states, the finite-dimensional density matrices can be obtain which is lead to closed-
form expressions for quantum information quantity. So, the finite-dimensional fermionic Dirac field
are more easily interpreted than the infinite-dimensional bosonic scalar field. TThis makes us eager to consider the behavior  of various concepts of quantum information on the
fermionic Unruh channel. In this work we will consider fermionic Unruh channel associated with Dirac field \cite{21}. It is observed that  in the fermionic channel the entanglement and teleportation fidelity  degrades by increasing the  acceleration. It is also observed that  Bell inequality does not violate for large value of acceleration. 
The work is organized as follow.  In Sec.\ref{sec2} we review the notion of the Unruh effect for a
two-mode fermionic system. In Sec.\ref{sec3}, we study the Unruh effect as a quantum noise channel. In Sec.\ref{sec4} we  review briefly the teleportation
fidelity, concurrence, and Bell-CHSH inequality. In Sec.\ref{sec5}, we provide some examples to study the effect of Unruh channel on the teleportation fidelity, concurrence, and Bell-CHSH inequality.  We summarize in Sec.\ref{sec6} the main results of this paper. 

\section{Unruh effect}\label{sec2}
The Unruh effect is particularly studied by exploring the Minkowski (flat) spacetime in terms of
Rindler coordinates. the space-time is divided into two disconnected region by  Rindler transformation such that, an accelerated observer in one region is separated
from the other region. Due to the fact that the limited field modes are not connected in these two different regions the quantum information of accelerated observer degrades and leads to thermal bath. As mentioned earlier, the fermionic field with few degree is considered in this work.  

Let us Consider the case in which the two observers Alice (A) and Rob (R) share a maximally entangled state of two Dirac field modes at Minkowski spacetime. In addition, we assume that each observer is equipped with detectors that are sensitive only to their respective modes. So, the quantum field is in a state
\begin{equation}\label{me}
\vert \psi \rangle_{AR} = \frac{1}{\sqrt{2}}\left(\left|0_{s}\right\rangle^{\mathcal{M}}\left|0_{k}\right\rangle^{\mathcal{M}}+\left|1_{s}\right\rangle^{\mathcal{M}}\left|1_{k}\right\rangle^{\mathcal{M}}\right),\end{equation}
where $\vert 0_j \rangle^{\mathcal{M}}$ and $\vert 1_j \rangle$ are vacuum and excitation states of the mode $j$ in Minkowski space. The state of Alice is created by the field mode $s$ and she has been equipped with detector sensitive to mode $s$ while Bob's state is constructed by the field mode $k$ and  he has been equipped with detector sensitive to mode $k$. When Bob moves with uniform acceleration $a$, the states corresponding to field mode $k$ must be redefined in Rindler bases in order to describe
what Rob sees. Considering the provided formalism the
Minkowski vacuum state converts to the Unruh mode, while the excited state is
a product state. In terms of modes in different
Rindler region $I$ and $II$, this concept can be expressed as
\begin{equation}\begin{array}{l}
\left|0_{k}\right\rangle^{\mathcal{M}}=\cos r\left|0_{k}\right\rangle_{I}\left|0_{k}\right\rangle_{I I}+\sin r\left|1_{k}\right\rangle_{I}\left|1_{k}\right\rangle_{I I} \\
\left|1_{k}\right\rangle^{\mathcal{M}}=\left|1_{k}\right\rangle_{I}\left|0_{k}\right\rangle_{I I}
\end{array}\end{equation}
where $\cos r=\frac{1}{\sqrt{1+e^{\frac{-2 \pi \omega c}{a}}}}$, $\omega$ is the Dirac particle frequency and $c$ is the speed of light in vacuum. Also worth mentioning that $a \in [0, \infty]$ and so we have $\cos r \in [\frac{1}{\sqrt{2}},1]$. Using this formulation the maximally entangled state in Eq.(\ref{me}) can be rewritten as
\begin{eqnarray}
\vert \psi \rangle &=&\frac{1}{\sqrt{2}}(\vert 0_s\rangle^{\mathcal{M}}(\cos r \vert 0_k \rangle_{I} \vert 0_k \rangle_{II}+ \\ \nonumber
&+&\cos r \vert 1_k \rangle_{I} \vert 1_k \rangle_{II})
+\vert 1_s\rangle^{\mathcal{M}}\vert 1_k \rangle_{I} \vert 0_k \rangle_{II} ),
\end{eqnarray}
Due to the fact that the two region $I$ and $II$ have no connection with each other in Rindler's space-time, it is possible to take a partial trace over zone $II$ and obtain the following density matrix
\begin{eqnarray}
\rho^{AI}&=&\frac{1}{2}[\cos^{2}r\vert 0_{s}^{\mathcal{M}},0_{k}^{I}\rangle \langle 0_{s}^{\mathcal{M}},0_{k}^{I}\vert  + \\ \nonumber
&+&  \cos r(\vert 0_{s}^{\mathcal{M}},0_{k}^{I}\rangle \langle 1_{s}^{\mathcal{M}},1_{k}^{I}\vert + \vert 1_{s}^{\mathcal{M}},1_{k}^{I}\rangle \langle 0_{s}^{\mathcal{M}},0_{k}^{I}\vert)+\\ \nonumber
&+& \sin^{2}r \vert 0_{s}^{\mathcal{M}},1_{k}^{I}\rangle \langle 0_{s}^{\mathcal{M}},1_{k}^{I}\vert +\vert 1_{s}^{\mathcal{M}},1_{k}^{I}\rangle \langle 1_{s}^{\mathcal{M}},1_{k}^{I}\vert].
\end{eqnarray}
In the following we will discuss about the channel interpretation of the Unruh effect.
\section{Characterization of Unruh Channel}\label{sec3}
In this section, Unruh channel is described from the insight of dynamical map in open quantum systems. Here the Choi-Jamiolkowski isomorphism is used to define  the Unruh channel. The Choi
matrix corresponding to Unruh channel $\varepsilon_U$, can be derived by applying the Unruh channel on one half of a maximally entangled two-qubit state as 
\begin{equation}
\rho_U=
  \left[ {\begin{array}{cccc}
   \cos^2 r & 0 & 0 & \cos r \\
   0 & \sin^2 r & 0 & 0 \\
   0 & 0 & 0 & 0 \\
   \cos r & 0 & 0 & 1 \\
  \end{array} } \right],
\end{equation}
The Kraus operators of the Unruh channel can be characterized by diagonalizing the Choi Matrix as \cite{23} 
\begin{equation}
 K_1=
  \left[ {\begin{array}{cc}
   \cos r & 0 \\
   0 & 1 \\
  \end{array} } \right],  \quad K_2=
  \left[ {\begin{array}{cc}
   0 & 0 \\
   \sin r & 0 \\
  \end{array} } \right],
\end{equation}
so, the Kraus representation of Unruh channel can be written as 
\begin{equation}
\varepsilon_U(\rho)=\sum_{j=1,2}K_j \rho K_j^{\dag}
\end{equation}
with the completeness condition
\begin{equation}
\sum_{j=1,2}K_j^{\dag} K_j = \mathcal{I}
\end{equation}
Looking at the Kraus operators of Unruh channel, it can be seen that this channel is similar to the amplitude damping channel represents the effect of a zero temperature thermal bath \cite{24,25}. However Unruh effect is associated with a finite temperature, so it is logical to expect that the Unruh channel correspond to the generalized amplitude damping channels.
\section{Teleportation, entanglement, and Bell
nonlocality in Unruh channel}\label{sec4}

In this paper, we consider the case in which inertial observer (Alice) wants to teleport to
accelerated observer (Rob) the one-qubit state
\begin{equation}
\vert \psi_{in} \rangle=\cos(\theta/2)\vert 0 \rangle + e^{i \phi}\sin(\theta/2)\vert 1 \rangle,
\end{equation}
where $\theta \in [0,\pi]$ and $\phi \in[0,2 \pi]$, while one can use a general two-qubit state $\rho  $ as the quantum channel for teleportation. In a standard teleportation protocol in which the Rob is allowed to perform any unitary transformation, wen reproducing the the teleported state, the maximum average teleportation fidelity will be given by \cite{Horodecki}
\begin{equation}
F_{av}(\rho)=\frac{1}{2}+\frac{1}{6}N(\rho  ),
\end{equation}
where $N(\rho  )=tr \sqrt{T^{\dag}T}$ and $T$ is a $3 \times 3$ positive matrix with the elements $T_{ij}=tr(\rho   \sigma_i \otimes \sigma_j)$ where $\sigma_{i,j}$'s are Pauli matrices. In order to teleport $\vert \psi_{in \rangle}$ with higher fidelity than purely classical communication
protocol, one requires the average fidelity is larger than $2/3$.

Since the existence of entanglement between Alice and Rob is necessary for teleportation we  consider the concurrence of $\rho  $ as 
\begin{equation}
C(\rho  )=\max \{0, \sqrt{\lambda_{1}}-\sqrt{\lambda_{2}}-\sqrt{\lambda_{3}}-\sqrt{\lambda_{4}}\},\end{equation}
where $\lambda_i$'s are the eigenvalues of the Hermitian operator $\rho\left(\sigma_{2} \otimes \sigma_{2}\right) \rho  ^{*}\left(\sigma_{2} \otimes \sigma_{2}\right)$ arranged in non-increasing order and $\rho  ^{*}$ represents the complex conjugate of $\rho  $. We will also study the Bell non-locality which is helpful to distinguish the two-qubit state $\rho  $ enabling non-classical teleportation
fidelity. For two-qubit states, the Bell non-locality can be detected by violation of the Bell-CHSH inequality  \cite{Clauser,Horodecki} 
\begin{equation}
\vert \langle B_{CHSH} \rangle_{\rho  } \vert = tr(\rho  B_{CHSH}) \leq 2,
\end{equation}
where $B_{CHSH}$ is the CHSH operator. The maximum of $\vert \langle B_{CHSH} \rangle_{\rho  } \vert$ can be find as 
\begin{equation}\label{max}
B_{max}(\rho)=2 \sqrt{u_1+u_2}
\end{equation}
 where $u_1$ and $u_2$ are two largest eigenvalues of the $T^{\dag}T$. the state $\rho  $ is Bell nonlocal if $B_{max}(\rho  )>2$. 
\section{Examples}\label{sec5}
In this section we consider some examples to study the effect of Unruh channel on  teleportation, entanglement, and Bell nonlocality. We will consider three different states with different features.  Alice and Rob first share the states which has mentioned in the examples. Then let Rob move away from stationary Alice with a uniform proper acceleration $a$. Finally we study the teleportation, entanglement, and Bell nonlocality of transformed state.
\subsection{Bell-diagonal state}
At first, we assume that Alice and Rob initially
being at inertial frame and share a Bell-diagonal state of two Dirac field modes
\begin{equation}\label{bell}
\rho^{AR}=\frac{1}{4}(I \otimes I + \sum_{i=1}^{3} r_{i}\sigma_i \otimes \sigma_i)
\end{equation}
where $\sigma_i$ ($i=1,2,3$) are Pauli matrices. The above density matrix is positive  if $\vec{r}=(r_1,r_2,r_3)$ belongs to a tetrahedron defined by the set of vertices $(-1,-1,-1)$,$(-1,1,1)$,$(1,-1,1)$ and $(1,1,-1)$. Let us the  case in which $r_1=1-2p$, $r_2=r_3=-p$. So, the state in Eq. (\ref{bell}) can be rewritten as
\begin{equation}
\rho^{AR}=p \vert \psi^- \rangle\langle \psi^- \vert + \frac{1-p}{2}(\vert \psi^+ \rangle\langle \psi^+ \vert + \vert \phi^+ \rangle\langle \phi^+ \vert),
\end{equation}
where $\vert \phi^{\pm} \rangle = \frac{1}{\sqrt{2}}[\vert 00 \rangle \pm \vert 11 \rangle]$ and $\vert \psi^{\pm} \rangle = \frac{1}{\sqrt{2}}[\vert 01 \rangle \pm \vert 10 \rangle]$ are the Bell diagonal states. Rob begins to move with an acceleration $a$ while Alice stay in inertial frame . This is equivalent to the effect of Unruh channel on Rob's state, so the transformed state can be obtained as 
\begin{equation}
\rho^{AI}=\sum_{i=1}^{2}(I \otimes K_i)\rho^{AR}(I \otimes K_{i}^{\dag}).
\end{equation}
So, the maximum average teleportation fidelity is obtained as
\begin{eqnarray}
F_{av}(\rho^{AI})&=&\frac{1}{2}+\frac{1}{6}\sqrt{p^{2}\cos^{4}r}+\frac{1}{12}\sqrt{\cos^{2}r(1-3p-(1-p)\cos r)^{2}} \nonumber \\
&+& \frac{1}{12}\sqrt{\cos^{2}r(3p-1-(1-p)\cos r)^{2}}.
\end{eqnarray}
The concurrence can be obtain as 
\begin{equation}
C(\rho^{AI})=2 \max \lbrace c_1, c_2 \rbrace
\end{equation}
where
\begin{eqnarray}
c_1&=&\frac{1}{4} \left(\left| (1-3 p) \cos r\right| -\frac{\sqrt{(p-1) \cos ^2 r ((p+1) \cos 2 r+p-3)}}{\sqrt{2}}\right),  \\
c_2&=&\frac{1}{4} \left(  \left| (p-1) \cos ^2 r \right| -\frac{\sqrt{(p+1) \cos ^2 r ((p-1) \cos 2 r+p+3)}}{\sqrt{2}} \right). \nonumber
\end{eqnarray}
In order to find the maximum of $\vert \langle B_{CHSH} \rangle_{\rho^{AR}  } \vert$ we have to find the eigenvalues of $T^{\dag} T$. These eigenvalues are
\begin{eqnarray}
u_1&=&\frac{1}{4} \cos ^2 r ((p-1) \cos r-3 p+1)^2, \nonumber \\
u_2&=&\frac{1}{4} \cos ^2 r ((p-1) \cos r+3 p-1)^2 \nonumber \\
u_3&=&p^2 \cos ^4 r.
\end{eqnarray}
So, from Eq.(\ref{max}) and considering the eigenvalues one can find the maximum of $\vert \langle B_{CHSH} \rangle_{\rho^{AR}  } \vert$ numerically. 

In Fig.\ref{figure1}, we show $F_{av}(\rho^{AI})$, $C(\rho^{AI})$ and $B_{max}(\rho^{AI})$ versus probability parameter for different value of acceleration $a$, when Alice and Rob initially share the Bell diagonal state. As can be seen from Fig.\ref{figure1}(a), concurrence decreases with increasing the acceleration of Rob. Fig.\ref{figure1}(b) shows that $F_{av}(\rho^{AI})$ decreases with increasing  the acceleration of Rob. As can be seen in inertial frame $a=0$ the teleportation fidelity is greater than $2/3$ for all value of probability parameter while for the case in which Rob moves with acceleration $a$ it is smaller than $2/3$ for some values of probability parameter. So, one can concluded that for these values of $p$ and $a$ the states are not good enough to support quantum teleportation protocol. From Fig.\ref{figure1}(c) it can be seen $B_{max}(\rho^{AI})$ decreases with increasing acceleration. It is observed that  the interval of $p$ for which Bell-CHSH inequality  is violated will be smaller with increasing acceleration.
\subsection{Werner state }
As a seconde example we consider the case in which Alice and Rob initially located in inertial frame and share a two-qubit werner state of two Dirac field modes
\begin{equation}
\rho^{AR}=\frac{1-p}{4}I \otimes I + p \vert \psi^{-} \rangle \langle \psi^{-} \vert,
\end{equation}
where $0 \leq p \leq 1$. Rob begins to move with an acceleration $a$ while Alice stay in inertial frame . This is equivalent to the effect of Unruh channel on Rob's state, so the transformed state can be obtained as 
\begin{equation}
\rho^{AI}=\left(
\begin{array}{cccc}
 \frac{1-p}{4} \cos ^2 r & 0 & 0 & 0 \\
 0 & \frac{1-p}{4}  \sin ^2 r+\frac{1+p}{4} & -\frac{p}{2}  \cos r & 0 \\
 0 & -\frac{p}{2}  \cos r & \frac{1+p}{4} \cos ^2 r & 0 \\
 0 & 0 & 0 & \frac{1+p}{4} \sin ^2 r+\frac{1-p}{4} \\
\end{array}
\right).
\end{equation}
The maximum average teleportation fidelity can be obtain as 
\begin{eqnarray}
F_{av}(\rho^{AI})&=&\frac{1}{2}+\frac{1}{6} p \cos ^2 r + \frac{p}{3} \cos r.
\end{eqnarray}
The concurrence becomes
\begin{equation}
C(\rho)=2 \max \lbrace 0, c_1\rbrace
\end{equation}
where 
\begin{equation}
c_1 = \left| -\frac{p}{2}  \cos r  \right| -\sqrt{\frac{1-p}{4}  \cos ^2(r) \left(\frac{1+p}{4} \sin ^2 r +\frac{1-p}{4}\right)}.
\end{equation}
In order to obtain the maximum of $\vert \langle B_{CHSH} \rangle_{\rho^{AR}  } \vert$ we have to find the eigenvalues of $T^{\dag} T$. These eigenvalues can be obtain as 
\begin{eqnarray}
u_1&=&p^{2}\cos^{2} r, \nonumber \\
u_2&=&p^{2}\cos^{2} r, \nonumber \\
u_3&=&2 p^{2} \cos^{4} r.
\end{eqnarray}
So, from Eq.(\ref{max}) and considering the eigenvalues one can find the maximum of $\vert \langle B_{CHSH} \rangle_{\rho^{AR}  } \vert$ numerically. 

In Fig.\ref{figure2}, $F_{av}(\rho^{AI})$, $C(\rho^{AI})$ and $B_{max}(\rho^{AI})$ are plotted interms of probability parameter $p$ for different value of acceleration $a$, when Alice and Rob initially share a two-qubit werner state. As can be seen from Fig.\ref{figure2}(a), concurrence decreases with increasing the acceleration of Rob. Fig.\ref{figure2}(b) shows that $F_{av}(\rho^{AI})$ decreases with increasing  the acceleration of Rob. It is observed that for both moving and inertial frame, $F_{av}(\rho^{AI})$ is smaller than $2/3$ for some values of probability parameter. So, one can concluded that for these values of $p$ and $a$ the states are not good enough to support quantum teleportation protocol. From Fig.\ref{figure2}(c) it can be seen $B_{max}(\rho^{AI})$ decreases with increasing acceleration. It is observed that  the interval of $p$ for which Bell-CHSH inequality  is violated will be smaller with increasing acceleration.
\subsection{X-state}
As an another example we consider the case in which Alice and Rob initially located in inertial frame and share a two-qubit X-state of two Dirac field modes
\begin{equation}
\rho^{AR}=(1-p) \vert \phi^{+} \rangle \langle \phi^{+} \vert + p \vert \psi^{+} \rangle \langle \psi^{+} \vert,
\end{equation}
where $0 \leq p \leq 1$. Rob begins to move with an acceleration $a$ while Alice stay in inertial frame . This is equivalent to the effect of Unruh channel on Rob's state, so the transformed state can be obtained as 
\begin{equation}
\rho^{AI}=\left(
\begin{array}{cccc}
 \frac{1-p}{2} \cos ^2 r  & 0 & 0 & \frac{1-p}{2}  \cos r \\
 0 & \frac{p}{2} +\frac{1-p}{2}  \sin ^2 r  & \frac{p}{2}  \cos  r & 0 \\
 0 & \frac{p}{2}  \cos r & \frac{p}{2}  \cos ^2 r & 0 \\
 \frac{1-p}{2}  \cos r & 0 & 0 & \frac{1-p}{2} + \frac{p}{2} \sin ^2 r \\
\end{array}
\right).
\end{equation}
The maximum average teleportation fidelity can be obtain as
\begin{eqnarray}
F_{av}(\rho^{AI})&=&\frac{1}{2}+\frac{1}{6} \cos r + \frac{2 p-1}{6} \cos r  + \frac{(2 p-1)^{2}}{6}\cos^{2} r.
\end{eqnarray}
To the maximum of $\vert \langle B_{CHSH} \rangle_{\rho^{AR}  } \vert$ we must find the eigenvalues of $T^{\dag} T$. These eigenvalues can be obtain as 
\begin{eqnarray}
u_1&=&\cos^{2} r, \nonumber \\
u_2&=&(1-2 p)^2 \cos ^2 r, \nonumber \\
u_3&=&(1-2 p)^2 \cos ^4 r.
\end{eqnarray}
So, from Eq.(\ref{max}) and considering the eigenvalues one can find the maximum of $\vert \langle B_{CHSH} \rangle_{\rho^{AR}  } \vert$ numerically. 

In Fig.\ref{figure3}, we plot $F_{av}(\rho^{AI})$, $C(\rho^{AI})$ and $B_{max}(\rho^{AI})$ as a function of probability parameter for different value of acceleration $a$, when Alice and Rob initially share the two-qubit X-state. As can be seen from Fig.\ref{figure3}(a), concurrence decreases with increasing the acceleration of Rob. Fig.\ref{figure3}(b) shows that $F_{av}(\rho^{AI})$ decreases with increasing  the acceleration of Rob. As can be seen in inertial frame $a=0$ the teleportation fidelity is greater than $2/3$ for all value of probability parameter while for the case in which Rob moves with acceleration $a$ it is smaller than $2/3$ for some values of probability parameter. So, one can concluded that for these values of $p$ and $a$ the states are not good enough to support quantum teleportation protocol. From Fig.\ref{figure3}(c) it can be seen $B_{max}(\rho^{AI})$ decreases with increasing acceleration. It is observed that  the interval of $p$ for which Bell-CHSH inequality  is violated will be smaller with increasing acceleration.
\section{Conclusion}\label{sec6}
In this work, the tools in quantum information were used to describe the Unruh-effect. The Unruh effect was considered as a quantum dynamical map with ordinary Kraus representation. In this work we studied the effect of Unruh channel on various quantum correlation such as  the quantum teleportation, entanglement, and Bell inequality violations for a Dirac field mode. We have shown that these correlations decrease with increasing observer acceleration in the moving frame. It has also shown that as a result of the Unruh effect, it is not possible to teleport optimally for some states.

%

%


%
%



\end{document}